%% file: 0_Main.tex
\documentclass[conference]{IEEEtran}
\IEEEoverridecommandlockouts
\usepackage{cite}
\usepackage{amsmath,amssymb,amsfonts}
\usepackage{algorithmic}
\usepackage{textcomp}
\usepackage{xcolor}
\usepackage{soul}
\usepackage{mathtools}
\usepackage{booktabs}
\usepackage{array}
\usepackage{enumitem}
\usepackage{float}
\usepackage{subfig}

\ifCLASSINFOpdf
  \usepackage[pdftex]{graphicx}
  \graphicspath{{.}{figures/}}
  \DeclareGraphicsExtensions{.pdf,.jpeg,.png}
\else
  \usepackage[dvips]{graphicx}
  \graphicspath{{../eps/}}
  \DeclareGraphicsExtensions{.eps}
\fi

\newcommand{\dx}[1]{\mathrm{d}#1}

\newcommand{\abs}[1]{\left|#1\right|}

\newcommand\scalemath[2]{\scalebox{#1}{\mbox{\ensuremath{\displaystyle #2}}}}

\allowdisplaybreaks

\makeatletter
\def\RemoveSpaces#1{\zap@space#1 \@empty}
\makeatother

\makeatletter
\renewcommand\st[1]{\@bsphack\@esphack}
\makeatother

\def\BibTeX{{\rm B\kern-.05em{\sc i\kern-.025em b}\kern-.08em
    T\kern-.1667em\lower.7ex\hbox{E}\kern-.125emX}}
\begin{document}

\title{Converging Radar and Communications in the Superposition Transmission \\
\thanks{This research was partly funded by the SESAR Joint Undertaking (SJU) in project NewSense (Evaluation of 5G Network and mmWave Radar Sensors to Enhance Surveillance of the Airport Surface), Grant Number 893917, within the framework of the European Union’s Horizon 2020 research and innovation program. The opinions expressed herein reﬂect the authors’ view only. Under no circumstances shall the SJU be responsible for any use that may be made of the information contained herein. This work was also partly supported by the Academy of Finland, under the project ULTRA (328226, 328214).}}

\author{Wenbo Wang\IEEEauthorrefmark{1}, Bo Tan\IEEEauthorrefmark{1}, Elena Simona Lohan\IEEEauthorrefmark{1}, Mikko Valkama\IEEEauthorrefmark{1},\\
\IEEEauthorblockA{\IEEEauthorrefmark{1}\textit{Faculty of Information Technology and Communication Sciences}, Tampere University, Finland
}
\{wenbo.wang, bo.tan, elena-simona.lohan, mikko.valkama\}@tuni.fi}

\maketitle

\begin{abstract}
This paper proposes a superposition transmission scheme for the future Radio Frequency (RF) convergence applications. The scheme is discussed under the assumption of a mono-static broadcasting channel topology. Under communications quality-of-service (QoS) constraints, the joint performance region of communications sum rate and radar estimation error variance is studied. Two radar signal waveforms, namely linear FM and parabolic FM, are used to investigate how signal shapes may influence the estimation accuracy. Both waveforms are generated with rectangular envelope. In the end, a numerical analysis is applied, which concludes that a moderate communications QoS promises a good communications fairness while with the limited radar performance degradation.   
\end{abstract}

\begin{IEEEkeywords}
joint radar and communications, superposition transmission, power domain, RF convergence, co-design
\end{IEEEkeywords}

\input{1_introduction}
\input{2_system_model}
\input{3_problem_formulation}

\input{4_performance_analysis}

\input{5_Conclusion}


\bibliographystyle{ieeetr}
\bibliography{bibliography/bib}

\end{document}

%% file: 1_introduction.tex
\section{Introduction} \label{sec:intro}




The booming wireless communication applications bring the need for more radio emitters and more spectrum resources, meanwhile causing a spectral congestion problem with legacy radar systems. At the same time, emerging applications, such as connected autonomous vehicles (CAV) and autonomous drones and robots, urge that the radio sensing and communications functions taking place in the common spectrum simultaneously. The above reasons drive the research on the convergence of two radio frequency (RF) systems when sensing and communication tasks will co-exist and be tackled in a joint manner. According to Bliss \textit{et al.} in \cite{Bliss_2017_TCCN} and \cite{Bliss_2017_Access}, the RF convergence can be categorised into three integration levels: \textbf{coexistence}, \textbf{cooperation} and \textbf{co-design}. In the coexistence level, radar and communication signal sources do not share any {\it a priori} information and consider the signal from the counter party as interference. In the cooperation level, a certain level of knowledge is shared between the radar and communications systems for a more effective interference cancellation. In the co-design level, the radar and the communication systems are designed from sketch for mutual/common benefits and by maximizing the use of spectral, time, and spatial resources.

In order to develop a highly integrated RF convergence system, the current research works often target to coordinate the signals in frequency, time, or spatial dimensions. The co-design in time domain can be traced back to 1960s, when pulse interval modulation (PIM) was proposed for embedded information on the radar pulses \cite{Mealey_1963_PIM}. In the frequency domain, the orthogonal frequency-division multiplexing (OFDM) waveform is often used for the dual-function design. In \cite{Sturm_2011_OFDMRadar}, authors demonstrated a vehicle detection function, which was implemented based on the OFDM communications signal. The recent research work in \cite{Taneli_2019_fullduplex} embraces the full-duplex circuit, which reduces the direct signal leakage and enables the detection of reflected Long-Term Evolution (LTE) and 5G New Radio (NR) OFDM signals from the drones and vehicles. In the spatial domain, multiple-input and multiple-output (MIMO), generalized to both phase coherent and spatial independent antenna arrays, is the main instrument to achieve the RF convergence. MIMO provides a high degrees-of-freedom (FoD) to differentiate and reduce the mutual interference between communications user and radar target by applying transmitting/receiving beamforming. MIMO configuration also achieves a high information rate, by leveraging the waveform diversity, and a high detection rate and resolution with a large physical aperture. \cite{Liu_2018_TWC} demonstrates a typical co-design based MIMO configuration, which leverages the null space of the communications for radar transmission.

In this paper, we envision a power-domain paradigm (called superposition transmission), in which the signal is fully superposed on frequency, spatial, and time domains for radar and communications co-design. To initiate the discussion, a mono-static broadcasting channel (MBC) topology \cite{Bliss_2017_Access}, which can be referred to downlink broadcasting communications channel, is used in this paper, as shown in Fig.~\ref{fig:sys_model}. This scenario is a typical downlink case according to 3GPP standards (3GPP TR36.859 \cite{3gpp_TR36.859)}). However, our case is different from the pure communications scenario in \cite{3gpp_TR36.859)}. The communications users 1 and 2 are also treated as radar targets (reflecting the radio signal) when receiving downlink data from the access node, and the transmitting power is split for achieving both communications and targets detection. This paper brings the following contributions to radar and communications co-design:
\begin{itemize}[leftmargin=*]
 \item A superposition waveform transmission method is tested for the first time, to the best of the Authors' knowledge, for radar and communications co-design. It releases the co-design from the constraint of the spectrum, space, and time orthogonality.
 \item Our work studies the impact of the quality of service (QoS) from communications' point of view on the performance of joint system.
 \item The proposed concept is verified in an MBC topology, which contains downlink communication channels, mono-static radar configuration, and entities (nodes) with mixed radar target and communications terminal, as shown in Fig.~\ref{fig:sys_model}. The topology fits future RF convergence applications, for example in the CAV and autonomous-drones scenarios.
\end{itemize}

The rest of the paper is organised as the follows Section~\ref{sec:sys_model} introduces the innovative setup of the superposition transmission of joint radar and communications system; Section~\ref{sec:pro_form} formulates the performance evaluation problem of the proposed dual-function system; Analytical analysis and Monte Carlo simulation are conducted and compared in Section~\ref{sec:per_analysis}; and the conclusions of the current work as well as a discussion about future works are given Section~\ref{sec:conclusion}.  


%% file: 2_system_model.tex
\section{System model} \label{sec:sys_model}
\begin{figure}[!t]
\centering 
\includegraphics[width=0.48\textwidth]{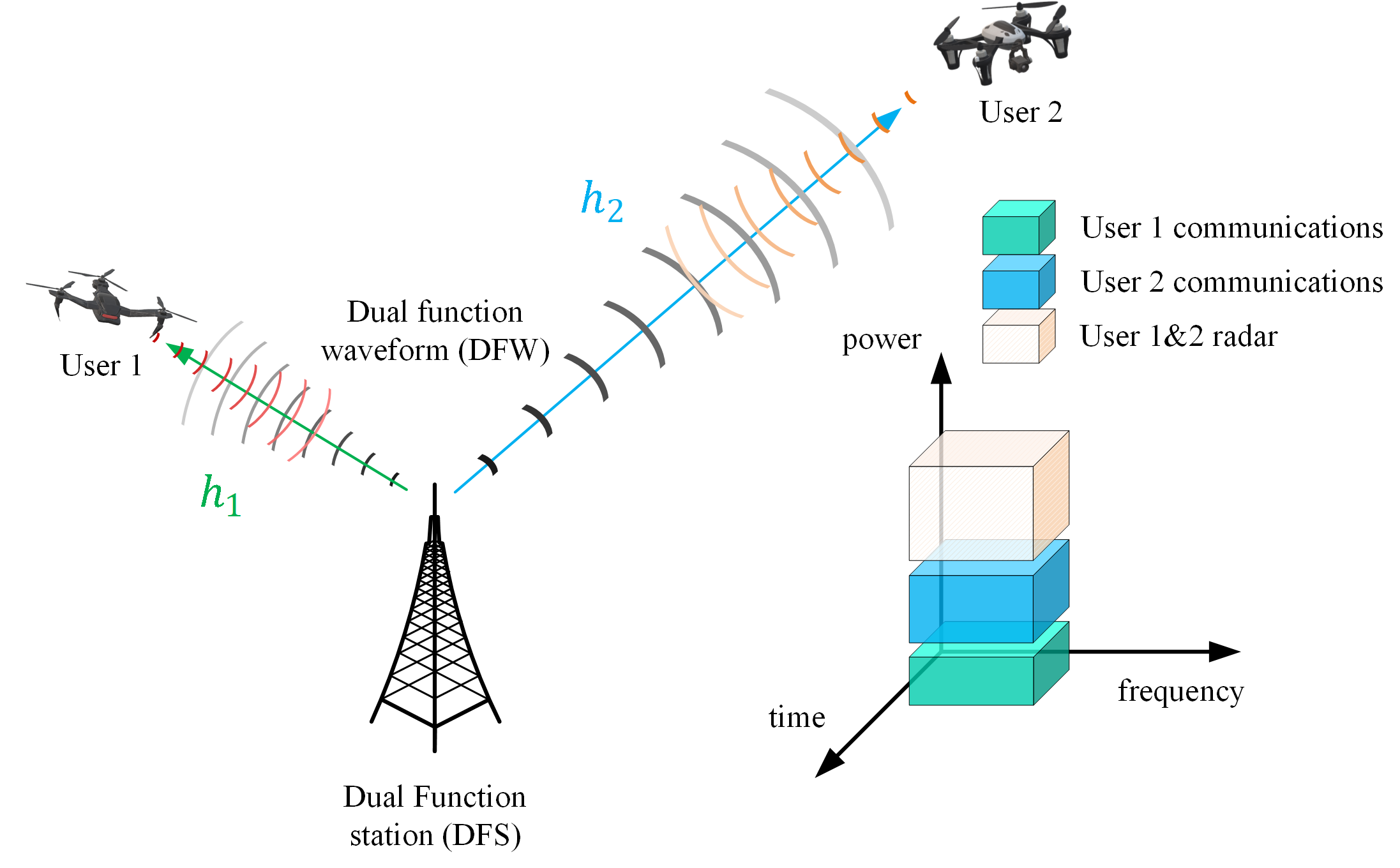}
\caption{The considered MBC topology and the illustration of resource block in our consideration.} \vspace{-0.5cm}
\label{fig:sys_model}
\end{figure}

Our purpose in this paper is to test a novel superposition-transmission-based RF convergence. Thus, we preclude spatial and spectral complexities by setting up an MBC topology scenario as shown in Fig.~\ref{fig:sys_model}, where all nodes are configured as the single carrier (SC) single-input single-output (SISO). In this MBC setup, there is one dual-function station (DFS) transmitting dual-function waveform (DFW) to user 1 and user 2 simultaneously, in the fully overlapped spectrum. We assume that both users are the communications nodes, meanwhile well-separated (not colinear with DFS and fall in different range bins) radar targets. The DFW is the superposing of downlink communications signals $s_1,s_2$ (for user 1 and 2 respectively) and radar signal $x$ (for both users), $\mathbb{E}(\left|s_1 \right|^2)=\mathbb{E}(\left|s_2 \right|^2)=\mathbb{E}(\left|x \right|^2)=1$. The channel gains for user 1 and user 2 are $h_1$ and $h_2$ \footnote{The block fading assumption is made in this paper, which means the fading process is approximately constant for a certain observation times, usually number of symbol intervals}. We use $\eta _1$ and $\eta _2$, for user 1 and user 2, to mimic the impacts of the radar cross-section (RCS) of users on the reflection signal strength. To be able to detect the reflected waveform from both users, the self-interference cancellation is conceived on the DFS. The recent experimental result in \cite{Taneli_2019_fullduplex} have proved that jointly applying of analog and digital cancellations can successfully weep of the self-interference and detect targets. In this paper, the residual self-interference is treated as attenuated instant transmitting signals by giving a coefficient $\xi$. The radar signal is a composition of repetition. We assume that the radar signal is known at all users and can be decoded then subtracted from the received superposition signal. In addition to above assumptions, we further attach two loose conditions: \textbf{\textit{i).}} the CSI is known at DFS and user terminal and \textbf{\textit{ii).}} the system works in a fast fading channel. These two conditions are not essential in this work; however, will facilitate the discussion.

Inspired by the Multi-user Superposition Transmission (MUST)\cite{3gpp_TR36.859)}, at the same time-and-frequency resource block, we propose a power-domain division scheme for joint radar and communications system as illustrated in Fig.~\ref{fig:sys_model}.

The total transmitted signals $S(t)$ are modeled by,
\begin{equation} \label{eq:tx_sig}
    S(t)=\alpha _1 s_1(t)+\alpha _2 s_2(t)+\alpha _r x(t)
\end{equation}
the power allocation coefficients for signals $s_1,s_2,x$ are $\alpha _1^2, \alpha _2^2, \alpha _r^2$ respectively, and $\alpha _1, \alpha _2, \alpha _r \in [0,1)$, without further specifications, we assume $\alpha _1 ^2,\alpha _2 ^2, \alpha _r ^2 \neq 0$ and $\alpha _1 ^2+\alpha _2 ^2+\alpha _r ^2 \leq 1$ for all the following analysis.

\subsection{Communications} \label{subsec:comm}
In the downlink, the received communications signals of user 1 $y_1(t)$ and user 2 $y_2(t)$ are respectively given by,
\begin{subequations} \label{eq:rx_comm}
    \begin{align}
        & y_1(t)=\abs{h_1}  S(t)+n_1(t)\\
        & y_2(t)=\abs{h_2}  S(t)+n_2(t)
    \end{align}
\end{subequations}
where $n_1(t) \sim \mathcal{N}(0,\sigma_1^2), n_2(t) \sim \mathcal{N}(0,\sigma_2^2)$ are additive white Gaussian noise (AWGN) at communications receivers of user 1 and user 2 respectively.

Since the superposition coding signals are transmitted, we follow the suggestion in \cite{tse2005fundamentals} that user 1 and 2 have disparate channels. If we assume $\left|h_1 \right|^2 > \left|h_2 \right|^2$, equivalently user 1 is the stronger user and user 2 is the weaker user, in received signal $y_1$ user 1 could first use SIC to detect signal $s_2$, then reconstruct signal by subtracting $s_2$. In $y_2$, under $\alpha _2 
^2 > \alpha _1^2$ the signal $s_2$ can be decoded while $s_1$ is treated as interference.

The Signal-to-Interference-plus-Noise-Ratio (SINR) for user 1 and 2, $\gamma_1$ and $\gamma_2$ are respectively expressed as,
\begin{subequations} \label{eq:sinr}
    \begin{align}
        & \gamma_1=\frac{\alpha _1^2\left|h_1 \right|^2}{\sigma _1 ^2} \label{eq:sinr_a}\\
        & \gamma_2=\frac{\alpha _2^2\left|h_2 \right|^2}{\left|h_2 \right|^2 \alpha _1^2 +\sigma _2 ^2} \label{eq:sinr_b}
    \end{align}
\end{subequations}

\subsection{Sensing} \label{subsec:sensing}
At the DFS, the echoes $z$ of the $k_{th}$ target (user) are modeled as,
\begin{equation} \label{eq:rx_radar}
    z_k(t) = \eta _k \abs{h_k}^2 S(t-\tau _k) + \xi S_{\text{int}} +n_r(t)
\end{equation}
where $\tau _k$ is the round-trip delay from the $k_{th}$ target, $n_r(t) \sim \mathcal{N}(0,\sigma _r^2)$ is AWGN. $\xi S_{\text{int}}$ is the self-interference residue and $\mathbb{E}(\abs{S_{\text{int}}}^2)=1$. According the recent works of the in-band self-cancellation \cite{Taneli_2019_fullduplex}\cite{bharadia2013full}, $105\sim110$ dB is an achievable self-interference suppression level which indicates that in-band residue is very close to the receiver sensitivity (noise floor) if 0 dBm power emission is set on the transmitting path. Thus, the residue term $\xi S_{\text{int}}$ is ignored in the following derivation. In the radar signal model, the backscatters from the clutters is another nagative impact fact which is assumed to be neutralized by the whitening filter before radar processing.

If the given radar task it to estimate the distance of the target (i.e., equivalently the time delay), an unbiased estimator has the minimum, which can be given by the Cram\'{e}r-Rao lower bound (CRLB). In mathematical form we have,
\begin{equation} \label{eq:crlb_exp}
    \mathbb{E}\big[ (\hat{\tau _k} - \tau _k)^2 \big] \geq \frac{1}{\mathbb{E}\bigg\{ \Big[ \frac{\partial}{\partial \tau _k} \log L\big(z(t);\tau _k  \big) \Big]^2 \bigg\}  }
\end{equation}
where $L\big(z(t);\tau _k  \big)$ is the likelihood function of $\tau _k$.

By considering the reflected communications components as the interference, we obtain the logarithm of likelihood function,
\begin{align} \label{eq:likelihood}
    &\log L\big(z(t);\tau _k  \big)  \\
    & =C -\frac{1}{2 \sigma _r^2} \Big(z(t)-\eta _k h_k^2 \alpha _k s_k  -\eta _k h_k^2 \alpha _r x (t-\tau _k)\Big)^2 \nonumber
\end{align}
where $C$ is a constant without involving $\tau _k$, hence the exact expression of $C$ is not provided here. The partial derivative yields to,
\begin{equation}
    \frac{\partial}{\partial \tau _k} \log L\big(z(t);\tau _k  \big) = -\frac{\eta _k h_k^2 \alpha _r}{\sigma _r^2} \cdot n_r(t) \cdot x'(t-\tau _k)
\end{equation}
straightforwardly,
\begin{equation} \label{eq:crlb_denominator}
    \mathbb{E}\bigg\{ \Big[ \frac{\partial}{\partial \tau _k} \log L\big(z(t);\tau _k  \big) \Big]^2 \bigg\} = \frac{\eta _k^2 h_k^4 \alpha _r^2}{\sigma _r^2} \mathbb{E} \Big\{\big[ x'(t-\tau _k) \big]^2 \Big\}
\end{equation}
Before further discussions, it is worth to mention the issue of handling the communications component in radar task. Communication components in the superposed waveform will also be reflected by the users and received by DFS. The reflected communications components can be used for enhancing the radar detection performance when it is adequately separated from the superposed waveform. Otherwise, the communications component will undermine the overall radar component (waveform) properties as the communications component is not well designed for detection purpose. To discuss the communication component for extra radar benefits will complicate our study as the benchmark of superposed waveform radar and communications co-design. Thus, in this work, the communication component is treated as interference for the detection task, the detection (radar) performance in this paper is a conservative estimation.

%% file: 3_problem_formulation.tex
\section{Problem formulation} \label{sec:pro_form}
The joint performance analysis is crucial in the evaluation of radar and communications co-design. In a communications system, we always put efforts to achieve maximum capacity (or sum-rate in multi-user scenario). In contrast, in radar systems, due to many shades of radar performance metrics, it is hard to determine the ultimate metric to evaluate radar systems comprehensively. In this section, we will discuss the formation of joint performance metrics hence the corresponding optimization problem.

\subsection{Universalizing the evaluation metric}
The communications rate of users 1 and 2 are denoted as $R_1$ and $R_2$, respectively, and they are upper bounded by,
\begin{subequations} \label{eq:comm_rate_bound}
    \begin{align}
    & R_1 \leq \log _2 (1+\gamma_1) \\
    & R_2 \leq \min \Big\{ \log _2 (1+\gamma_2), \log _2 (1+\overline{\gamma}_2) \Big\}
    \end{align}
\end{subequations}
where $\log _2 (1+\overline{\gamma}_2)$ is the upper-bound communication rate for SIC on user 1 to successfully decode $s_2$, $\overline{\gamma}_2$ is given by,
\begin{equation}
    \overline{\gamma}_2=\frac{\alpha _2^2\left|h_1 \right|^2}{\left|h_1 \right|^2\alpha _1^2 +\sigma _c ^2}
\end{equation}

Provided the channel gains assumption, the communication sum rate $R_{\text{sum}}$ of users 1 and 2 is upper bounded by,
\begin{equation} \label{eq:sum_rate}
    R_{\text{sum}} \leq \smashoperator{\sum _{k=1} ^2} \log _2 (1+\gamma_k)
\end{equation}

In the radar system, as we mentioned in Section~\ref{subsec:sensing}, the CRLB is a popular metric used to evaluate the parameters estimation, which implies the performance of system. For time delay estimation \cite{cook2012radar}, given \eqref{eq:crlb_exp} and \eqref{eq:crlb_denominator} we can have,
\begin{equation}
    \mathbb{E} \Big\{\big[ x'(t-\tau _k) \big]^2 \Big\} = 2EWB_{\mathrm{rms}} ^2
\end{equation}
where $W$ is the bandwidth, $2E$ is the total received signal energy and $B_{\mathrm{rms}}$ is the root-mean-square (rms) bandwidth, given by,
\begin{subequations}
\setlength{\abovedisplayskip}{3pt}
\setlength{\belowdisplayskip}{3pt}
\begin{equation} \label{eq:energy}
    2E = \smashoperator{\int _0 ^T} \abs{x(t)}^2 \dx{t}
\end{equation}
\begin{equation} \label{eq:rms_band}
    B_{\mathrm{rms}} ^2 = \frac{4\pi ^2 \smashoperator{\int _{- \infty} ^{\infty}} f^2\abs{X(f)}^2 \dx{f}}{\smashoperator{\int _{- \infty} ^{\infty}} \abs{X(f)}^2 \dx{f}}
\end{equation}
\end{subequations}
where $T$ is the radar receiving duration, $X(f)$ is the spectrum of signal $x(t)$. Hence,
\begin{equation}
    \mathbb{E}\big[ (\hat{\tau _k} - \tau _k)^2 \big] \geq \frac{\sigma _r^2}{2\eta _k^2 h_k^4 \alpha _r^2 EWB_{\mathrm{rms}} ^2 }
\end{equation}
clearly $\frac{2EW}{\sigma _r^2}$ is the radar receiving Signal-to-Noise-Ratio (SNR). Until here, the total estimation error variance $\sigma _{\epsilon}^2$ can be derived as,
\begin{equation} \label{eq:est_error}
    \sigma _{\epsilon}^2 \geq \smashoperator{\sum _{k=1} ^2} \frac{\sigma _r^2}{2\eta _k^2 h_k^4 \alpha _r^2 EWB_{\mathrm{rms}} ^2 }
\end{equation}

\subsection{Optimization problems}
The communications sum rate and the estimation error variance of radar system do not share the same unit of measurements. We can only consider the radar and communications co-design as two sub-problems,
\begin{equation} \label{eq:opt_comm} 
\centering
\begin{array}{cl}
\displaystyle \smashoperator{\max _{\alpha _1, \alpha _2}} & R_{\text{sum}} \\
\text{s.t.} & \alpha _1 ^2+\alpha _2 ^2  \leq 1 - \alpha _r ^2, \\
            & R_2 \geq R_{0,2}
\end{array}
\end{equation}
\begin{equation} \label{eq:opt_radar} 
\centering
\begin{array}{cl}
\displaystyle \smashoperator{\min _{\alpha _1, \alpha _2, \alpha _r}} &  \sigma _{\epsilon}^2 \\
\text{s.t.} & \alpha _1 ^2+\alpha _2 ^2  \leq 1 - \alpha _r ^2, \\
            & R_2 \geq R_{0,2}, R_1 \geq R_{0,1}
\end{array}
\end{equation}
where $R_{0,2},R_{0,1}$ are the minimum rate to guarantee the QoS for user 2 and user 1 respectively. $\alpha _r^2$ is treated as a parameter in \eqref{eq:opt_comm}, given a certain tolerance of radar estimation error, we would like to achieve the sum rate maximum.


%% file: 4_performance_analysis.tex
\section{Performance analysis} \label{sec:per_analysis}
We derive the joint performance boundaries and the corresponding power allocations for communications users and radar function.

\subsection{Optimal solutions} \label{subsec:op_sol}
We first determine the feasible region for $\alpha_1^2,\alpha_2^2,\alpha_r^2$ from problem \eqref{eq:opt_comm}, then find some feasible points in problem \eqref{eq:opt_radar}.
\subsubsection{Communications rate}
in problem \eqref{eq:opt_comm}, $R_{\text{sum}}$ is a binary logarithm function of the product between $1+\gamma_1$ and $1+\gamma_2$, and it will monotonically increase with this product. To find the optimum in \eqref{eq:opt_comm}, provided the constraint $R_2 \geq R_{0,2}$, it is of interest to investigate the monotonicity of the above product as a function $f_1$,
\begin{equation}
    f_1 = (1+\gamma_1) (1+\gamma_2)
\end{equation}
If we consider $f_1$ as the function of $\alpha _2^2$, the following form is obtained,
\begin{equation} \label{eq:func_opt_comm}
\scalemath{0.9}
{
    \medmuskip = 1mu
    \thinmuskip = 0mu
    f_1(\alpha _2^2)=1+\frac{\alpha _1^4\left|h_1 \right|^2\left|h_2 \right|^2+\alpha _1^2\left|h_1 \right|^2 \sigma _2^2 + \alpha _2^2\left|h_2 \right|^2 \sigma _1^2+\alpha _1^2\alpha _2^2\left|h_1 \right|^2\left|h_2 \right|^2}{\alpha _1^2 \left|h_2 \right|^2 \sigma _1^2 + \sigma _1^2\sigma _2^2}
}
\end{equation}
and the first order partial derivative of \eqref{eq:func_opt_comm} is,
\begin{equation} \label{eq:deri_func_opt_comm}
    \frac{\partial f_1}{\partial \alpha _2^2}=-\frac{(\left|h_1 \right|^2 \sigma _2^2 - \left|h_2 \right|^2 \sigma _1^2)(\left|h_2 \right|^2 \kappa + \sigma _2^2)}{\big[ \left|h_2 \right|^2 (\kappa - \alpha _2^2)+\sigma _2^2 \big]^2 \sigma _1^2}
\end{equation}
where $\kappa=1-\alpha _r^2$ and $\alpha _1^2+\alpha _2^2=\kappa$. Since our assumption is $\left|h_1 \right|^2 > \left|h_2 \right|^2$, under $\sigma _1^2 \leq \sigma _2^2$ $^($\footnote{in this paper, we will not provide discussions for the situation where $\sigma _1^2$ is much greater than $\sigma _2^2$.}$^)$ the $\frac{\partial f_1}{\partial \alpha _2^2}$ will always be negative for all the feasible values of $\alpha _2^2$. Consequently $R_{\text{sum}}$ in \eqref{eq:opt_comm} will monotonically decrease with $\alpha _2^2$ increasing, and reaches its maximum when $R_2=R_{0,2}$ holds. Now we have,
\begin{equation} \label{eq:eq_opt_comm}
    \log _2(1+\frac{\alpha _2^2\left|h_2 \right|^2}{\left|h_2 \right|^2 (\kappa - \alpha _2^2) +\sigma _2 ^2})=R_{0,2}
\end{equation}
the solutions of \eqref{eq:eq_opt_comm} are,

\begin{subequations} \label{eq:power_alloc_sol}
\begin{align}
    & \alpha _1 ^2 =\frac{\kappa\left|h_2 \right|^2-\sigma _2^2(2^{R_{0,2}}-1)}{\left|h_2 \right|^2 2^{R_{0,2}}} \label{eq:power_alloc_sol_1} \\
    & \alpha _2 ^2 = \frac{(2^{R_{0,2}}-1)(\kappa\left|h_2 \right|^2+\sigma _2^2)}{\left|h_2 \right|^2 2^{R_{0,2}}}
\end{align}
\end{subequations}
under the solution \eqref{eq:power_alloc_sol}, $R_{\text{sum}}$ achieves its optimum. The results in \eqref{eq:power_alloc_sol} are also consistent with \cite{wang2016power}. 

\subsubsection{Radar estimation error}
in problem \eqref{eq:opt_radar}, the minimum of $\sigma _{\epsilon}^2$ monotonically decreases with $\alpha _r$, the larger value of $\alpha _r^2$ we give the smaller $\sigma _{\epsilon}^2$ we have.

Under constraint $R_2 \geq R_{0,2}, R_1 \geq R_{0,1}$, we could have,
\begin{equation} \label{eq:comm_power_constraint}
    \left\{
    \begin{aligned}
    & \alpha _1^2 \geq \frac{(2^{R_{0,1}}-1)\sigma_1^2}{\left|h_1 \right|^2} \\
    & \alpha _2^2 \geq (2^{R_{0,2}}-1) \big[ \frac{(2^{R_{0,1}}-1)\sigma _1^2}{\left|h_1 \right|^2 }+\frac{\sigma _2^2}{\left|h_2 \right|^2} \big]
    \end{aligned}
    \right.
\end{equation}
the maximum of $\alpha _r^2$ is achieved when $\alpha _1^2,\alpha _2^2$ in \eqref{eq:comm_power_constraint} are at minimum values. Based on \cite{cook2012radar}, we present the energy $E$ and rms bandwidth $B_{\mathrm{rms}}$ under two modulations of radar waveform: linear FM with rectangular envelope and parabolic FM with rectangular envelope.
\subsubsection{Linear FM with rectangular envelope} for a linear frequency modulation (FM) signal with $\mathrm{rect}(t/T)$ envelope, energy $E_{\text{linear}}$ and rms bandwidth $B_{\mathrm{rms}}^{\text{linear}}$ yield to,
\begin{subequations} \label{eq:E_B_linear_rect}
\setlength{\abovedisplayskip}{3pt}
\setlength{\belowdisplayskip}{3pt}
\begin{equation}
    E_{\text{linear}}=\frac{T}{2}
\end{equation}
\begin{equation}
    B_{\mathrm{rms}}^{\text{linear}}=\frac{\pi ^2 W^2}{3}
\end{equation}
\end{subequations}
hence \eqref{eq:est_error} alters to,
\begin{equation}
    \sigma _{\epsilon, \text{linear}}^2 \geq \smashoperator{\sum _{k=1} ^2} \frac{3\sigma _r^2}{\pi ^2 \eta _k^2 h_k^4 \alpha _r^2 TW ^3 }
\end{equation}

\subsubsection{Parabolic FM with rectangular envelope} for a parabolic FM signal with rectangular envelope, energy $E_{\text{parabolic}}$ and rms bandwidth $B_{\mathrm{rms}}^{\text{parabolic}}$ yield to,
\begin{subequations}
\setlength{\abovedisplayskip}{3pt}
\setlength{\belowdisplayskip}{3pt}
\begin{equation}
    E_{\text{parabolic}}=\frac{T}{2}
\end{equation}
\begin{equation}
    B_{\mathrm{rms}}^{\text{parabolic}}=\frac{16\pi ^2 W^2}{45}
\end{equation}
\end{subequations}
hence \eqref{eq:est_error} alters to,
\begin{equation}
    \sigma _{\epsilon, \text{parabolic}}^2 \geq \smashoperator{\sum _{k=1} ^2} \frac{45\sigma _r^2}{16 \pi ^2 \eta _k^2 h_k^4 \alpha _r^2 TW ^3 }
\end{equation}

\subsection{Numerical results}
We implement a series of simulations to validate our theoretical analysis and to present joint performance in the radar and communications co-design explicitly. Table~\ref{tab:para} lists the parameters in the simulations.
\begin{table}[!htbp]
\setlength{\tabcolsep}{4pt}
\caption{Parameters in simulations}
\centering
\begin{tabular}{p{2.5cm}m{1.3cm}p{2.5cm}m{1.3cm}} 
\toprule
\textbf{Parameters} &  \textbf{Value} & \textbf{Parameters} &  \textbf{Value}  \\ \midrule
channel gain \footnotemark $\left|h_1 \right|^2$ & $-90$ dB & channel gain $\left|h_2 \right|^2$ & $-100$ dB \\ \midrule
noise in user 1 $\sigma_1^2$ & $-105$ dBm & noise in user 2 $\sigma_2^2$ & $-105$ dBm \\\midrule
noise in radar $\sigma_r^2$ & $-110$ dBm & self-interference cancellation & $110$ dB \\ \midrule
user 1 RCS $\eta_1$ & $0.1$ $\text{m}^2$ & user 2 RCS $\eta_2$ & $0.5$ $\text{m}^2$\\
\midrule
bandwidth $W$ & $20$ $\text{MHz}$ & Time-bandwidth product $TW$ & $1000$\\
\bottomrule
\end{tabular} \vspace{-0.3cm}
\label{tab:para}
\end{table}

In the optimization problem \eqref{eq:opt_comm} and \eqref{eq:opt_radar}, we expect to observe two behaviours of co-design system,
\begin{itemize}
    \item Given the weak user QoS $R_{0,2}$ in \eqref{eq:opt_comm}, the trend of $R_{\text{sum}}$ with $\sigma _{\epsilon}^2$;
    \item Given respectively the weak and strong user QoS $R_{0,2},R_{0,1}$ in \eqref{eq:opt_radar}, the minimum of $\sigma _{\epsilon}^2$.
\end{itemize}

\footnotetext{the path loss is incorporated into the channel gain.}

\begin{figure}[!htbp]
\subfloat[Linear FM with rectangular envelope]{\includegraphics[width=0.45\textwidth]{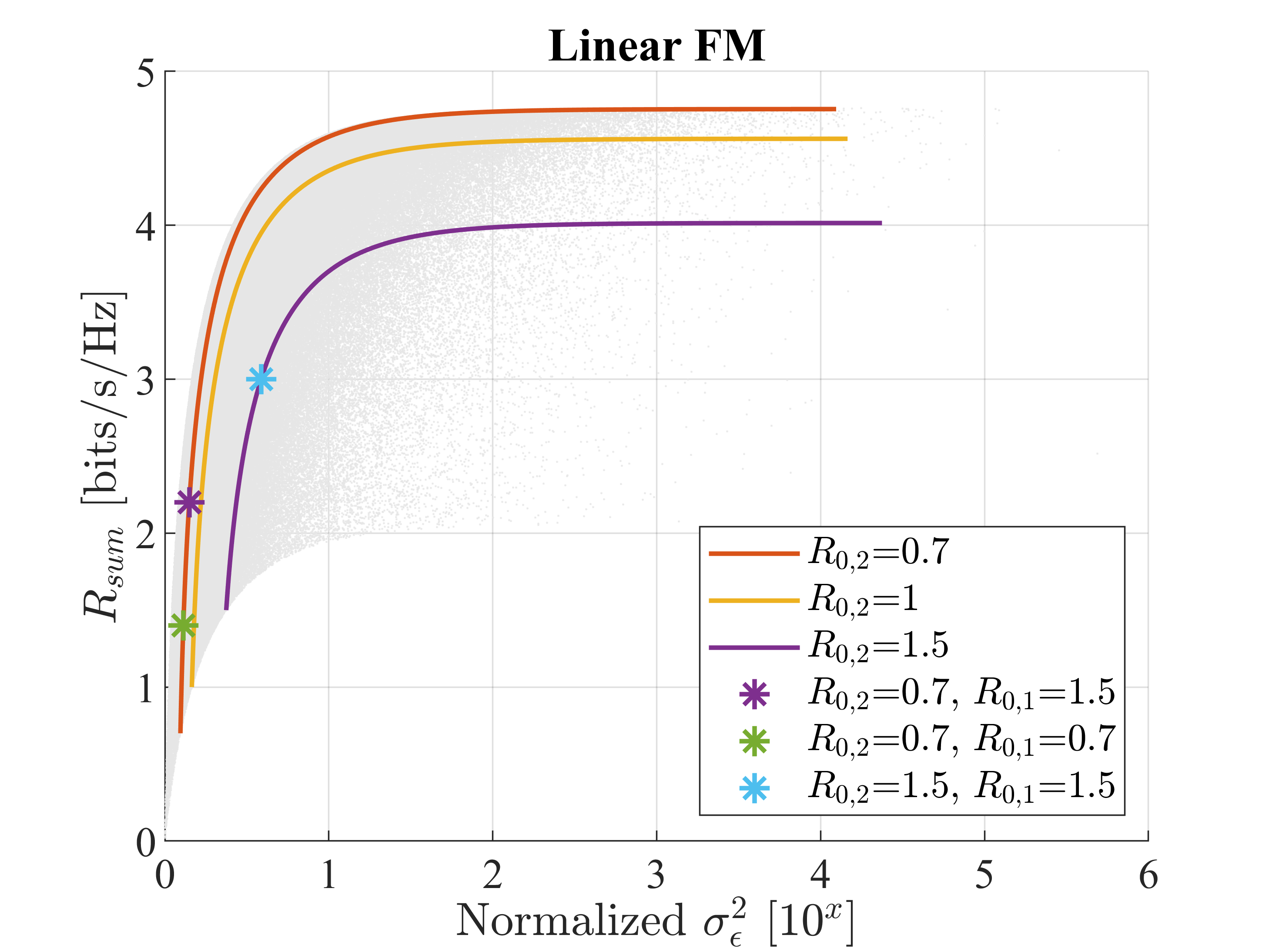} \label{subfig:linear_FM}} \\ 
\subfloat[Parabolic FM with rectangular envelope]{\includegraphics[width=0.45\textwidth]{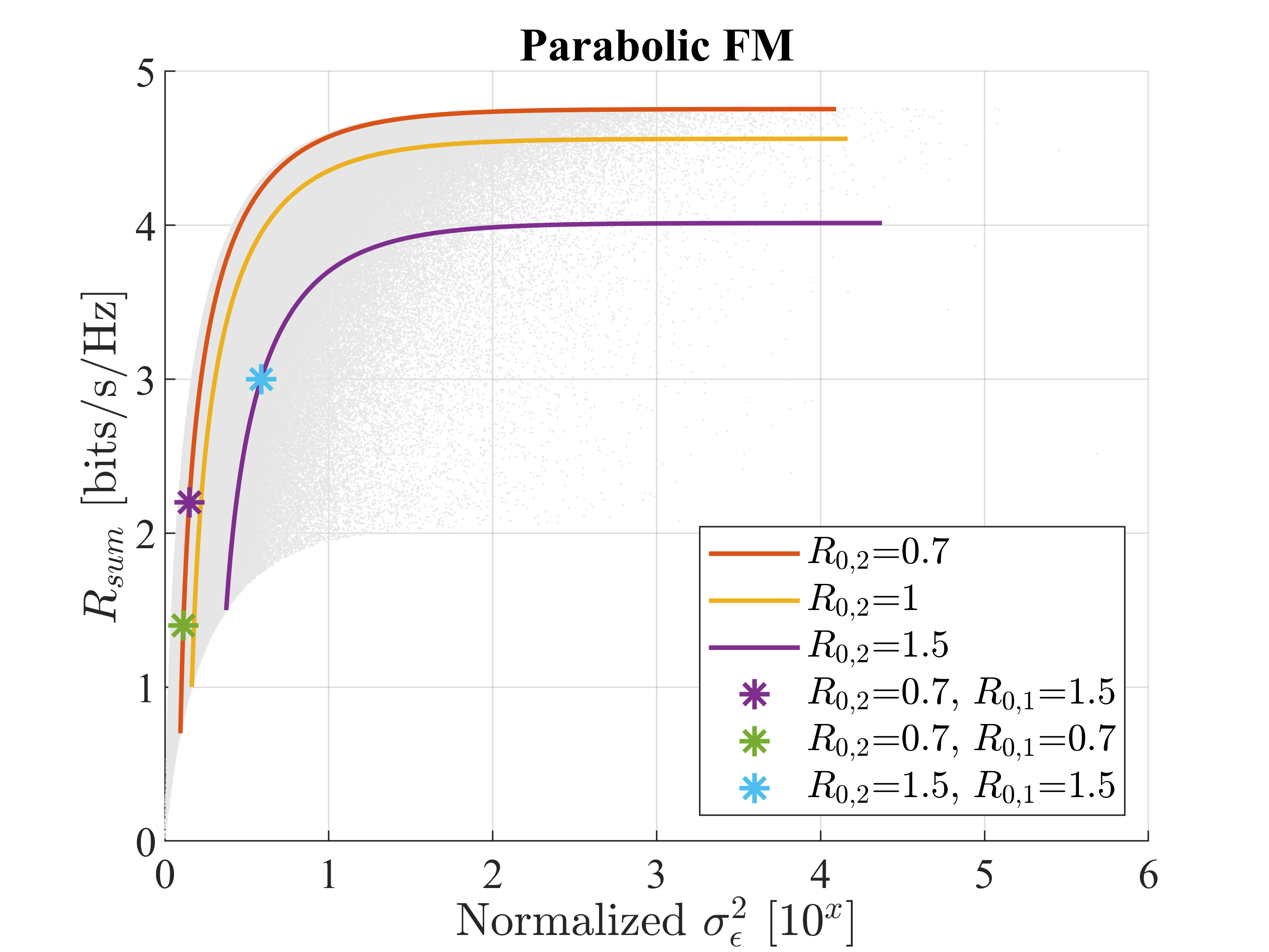} \label{subfig:parabolic_FM}} 
\caption{$R_{\text{sum}}$ Versus normalized $\sigma _{\epsilon}^2$ under linear FM and parabolic FM. Normalized $\sigma _{\epsilon}^2$ is given by the power of $10$ to better scale up.} \vspace{-0.5cm}
\label{fig:FM}
\end{figure}

Fig.~\ref{fig:FM} presents the numerical results of optimization problems \eqref{eq:opt_comm} and \eqref{eq:opt_radar}. Six cases are discussed in the figure, $R_{\text{sum}}$ versus $\sigma _{\epsilon}^2$ under $\smash{R_{0,2}=0.7,1,1.5}$ (unit:bits/s/Hz); minimum $\sigma _{\epsilon}^2$ under $\smash{\{R_{0,2}=0.7,R_{0,1}=1.5\}}$, $\smash{\{R_{0,2}=0.7,R_{0,1}=0.7\}}$, $\smash{\{R_{0,2}=1.5,R_{0,1}=1.5\}}$. The solid lines indicate the relationship between $R_{\text{sum}}$ and $\sigma _{\epsilon}^2$ under constraint $R_2 \geq R_{0,2}$. All star (i.e., *) markers show the minimum $\sigma _{\epsilon}^2$ under constraints $R_2 \geq R_{0,2}, R_1 \geq R_{0,1}$. The grey dots background gives the feasible region of the relationship between $R_{\text{sum}}$ and $\sigma _{\epsilon}^2$. $\sigma _{\epsilon}^2$ is normalized by the minimum of $\sigma _{\epsilon}^2$ in the figures,
\begin{equation}
    \min(\sigma _{\epsilon}^2) = \sigma _{\epsilon}^2 |_{\alpha _r^2=1}
\end{equation} \vspace{-0.3cm}

With the increase of QoS requirement in user 2 (the weak user), the radar performance drops rapidly. Comparing with user 2, the QoS requirement of user 1 (the strong user) has less influence in the degrades of radar estimation accuracy. As we can see from Fig.~\ref{fig:FM}, horizontally the distance between purple star marker (i.e., $\smash{R_{0,2}=0.7,R_{0,1}=1.5}$) and blue star marker (i.e., $\smash{R_{0,2}=1.5,R_{0,1}=1.5}$) is much larger than that between purple star marker (i.e., $\smash{R_{0,2}=0.7,R_{0,1}=1.5}$) and green star marker (i.e., $\smash{R_{0,2}=0.7,R_{0,1}=0.7}$). Observing the solid lines in the figure, up to a point $R_{\text{sum}}$ tends to converge no matter how much power we allocate to communications signals. This phenomenon also implies that the unreasonable large QoS requirement in the weak user could jeopardize the whole co-design system. By comparing Fig.~\ref{subfig:linear_FM} with Fig.~\ref{subfig:parabolic_FM}, we may conclude that under rectangular envelope linear FM and parabolic FM barely show difference on the performance of co-design system.

In the superposition transmission scheme, a moderate QoS requirement for the weak user (e.g., the user with lower channel gain) could benefit both communications and sensing. However, the performance of radar system abruptly drops when the QoS requirements of weak users increase over a certain threshold; in other words, from Fig.~\ref{fig:FM} the horizontal distance between the solid purple line (i.e., \smash{$R_{0,2}=1.5$}) and the solid orange line (i.e., \smash{$R_{0,2}=1$}) is much larger than the horizontal distance between the solid orange line (i.e., \smash{$R_{0,2}=1$}) and the solid red line the solid orange line (i.e., \smash{$R_{0,2}=0.7$}).

For a communications-priority system, in both Fig.~\ref{subfig:linear_FM} and Fig.~\ref{subfig:parabolic_FM} the upper-left corner of solid lines would be the optimal operation point, due to the convergence of sum rate for communications. At these points, the minimum radar estimation error variance also reaches after maximum achievable sum rate has been met.

To further reveal the communications system performance in the co-design system, we adopt Jain's fairness,
\begin{equation} \label{eq:jain_fairness}
    \mathcal{J}(x_1,x_2,\dots,x_n)=(\displaystyle \smashoperator{\sum _{i=1}^n} x_i )^2/ \displaystyle \smashoperator{\sum _{i=1}^n} x_i^2
\end{equation}

\begin{figure}[!htbp]
\centering
\includegraphics[width=0.45\textwidth]{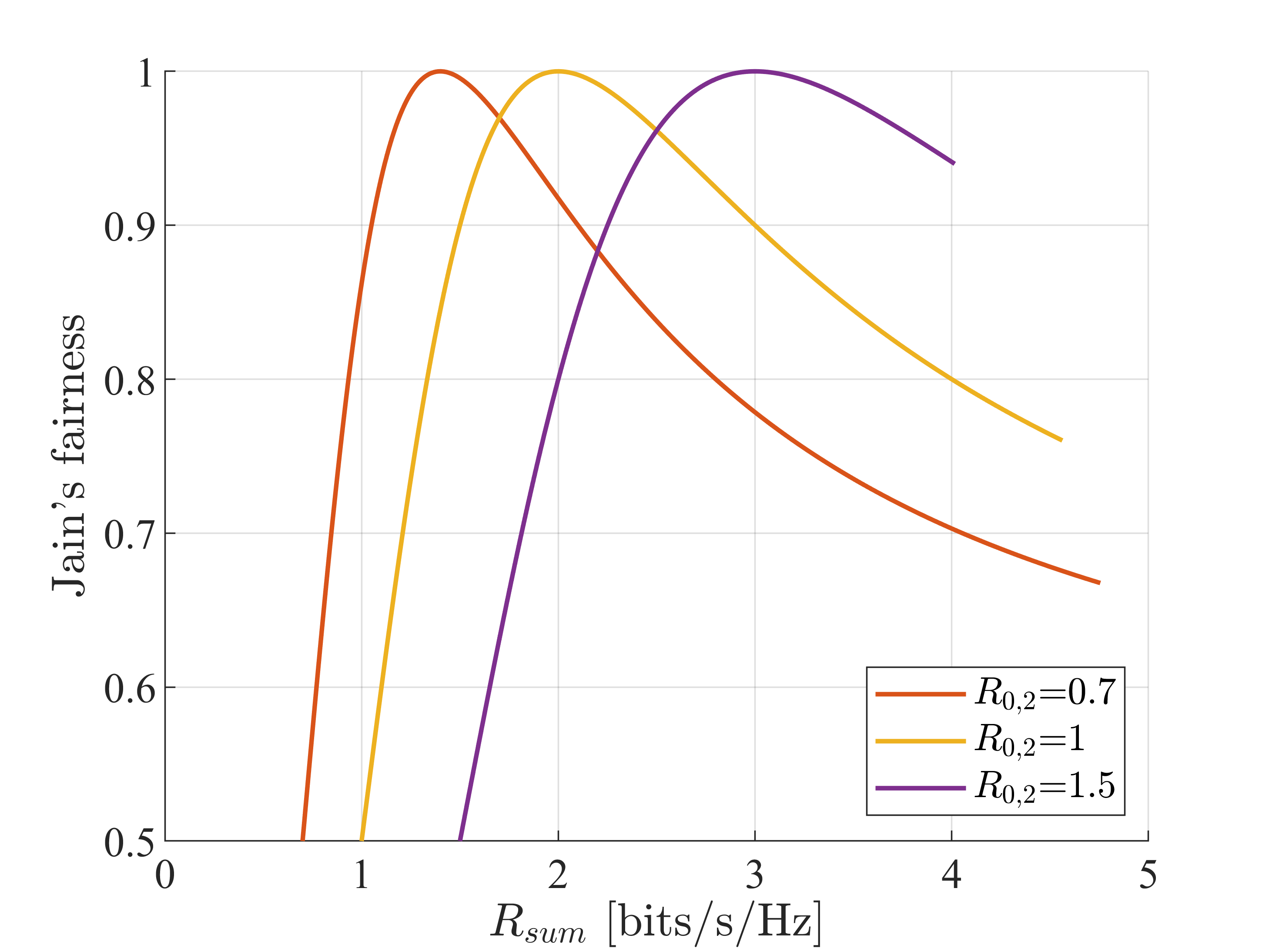}
\caption{The Jain's fairness between two users in communications, under various minimum QoS requirements of user 2.} \vspace{-0.3cm}
\label{fig:fairness}
\end{figure}

Fig.~\ref{fig:fairness} illustrates the fairness versus $R_{\text{sum}}$ under different QoS requirement in weak user. Clearly, low QoS requirement leads to low fairness during most of $R_{\text{sum}}$ values; high QoS case behaves even worse than low QoS case in the fairness, the highest reachable $R_{\text{sum}}$ values is relatively far from the maximum. A moderate QoS requirement guarantees both high fairness and high $R_{\text{sum}}$ value.

As a result, a moderate QoS requirement for the weak user promises the good fairness of communications and low estimation errors of sensing.

\begin{figure}[!htbp]
\centering
\includegraphics[width=0.45\textwidth]{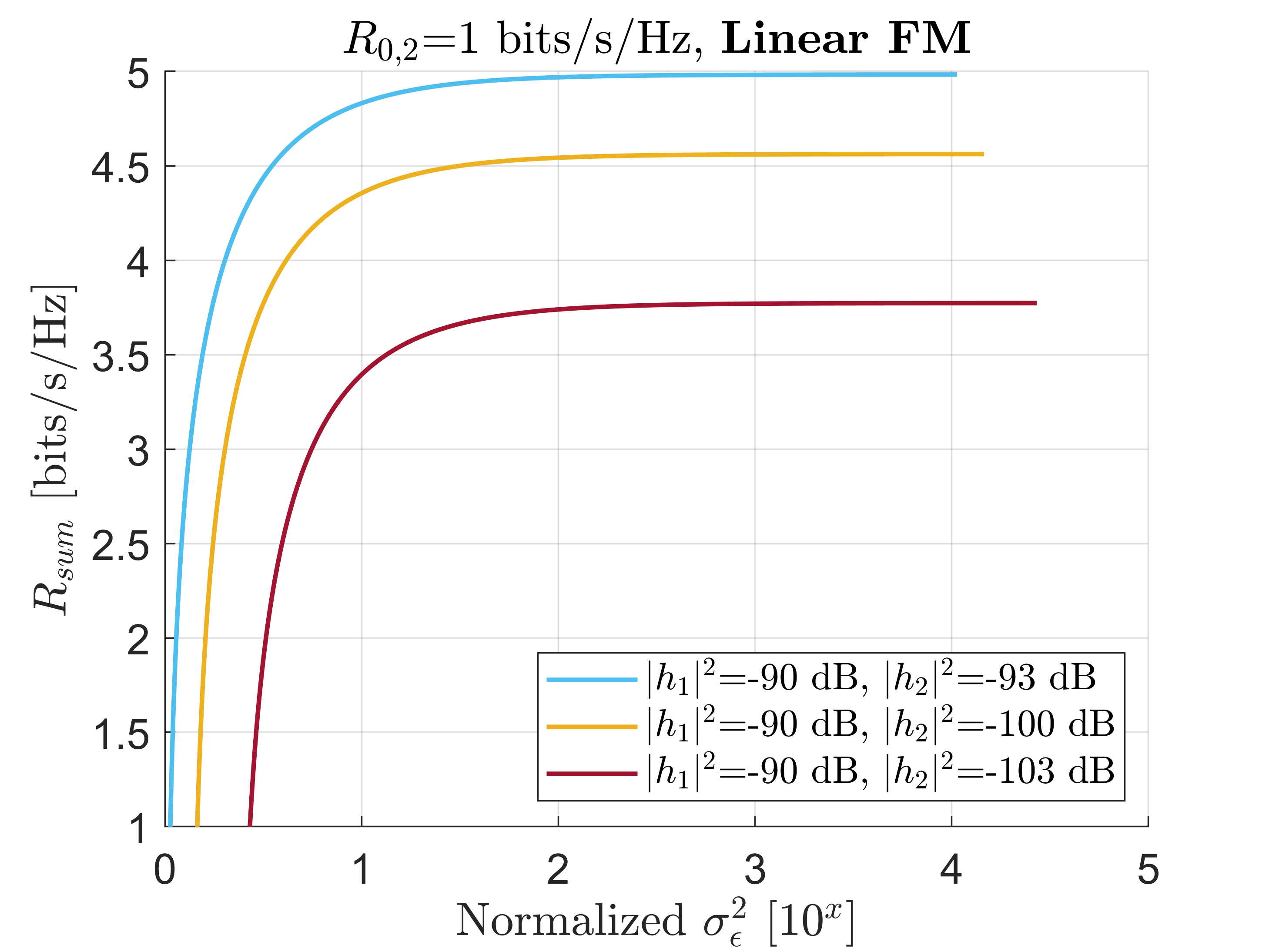}
\caption{$R_{\text{sum}}$ Versus normalized $\sigma _{\epsilon}^2$ under different level of asymmetry between two users' channel.} \vspace{-0.2cm}
\label{fig:linear_FM_asy}
\end{figure}

Fig.~\ref{fig:linear_FM_asy} demonstrates how the asymmetry of users' channel could affect the system performance. This numerical analysis shows that the asymmetry between two users' channel has impact on the system performance, the greater the level of asymmetry is, the larger degradation of the system is. The above result implies the impact of user grouping strategies on the overall system performance, which is that low level of asymmetry between users' channel compromises less on radar performance.

%% file: 5_Conclusion.tex
\section{Conclusion} \label{sec:conclusion}
In this paper, the superposition transmission is proposed for radar and communication co-design in contrast with the conventional frequency, time, and spatial domain operations. The joint radar-and-communications performance is analysed in the mono-static broadcasting topology. A generic communications signal is utilized together with two specific radar signals, namely a linear FM and a parabolic FM radar signal. A moderate QoS requirement for a weak user balances both communications fairness and the overall system performance. Low level of asymmetry between users' channel implies better co-design system performance. Conservatively speaking, in the joint radar-and-communication system, the radar side has large demands on the power allocation, which leads to the low-to-moderate communications rates. This superposition transmission scheme, under current study, is not suitable for high data rate applications; whereas it is a good scheme for drones control and command signals, together with sensing functionality.

The current findings in this work can be seen  as a bedrock for future extensions on \textbf{\textit{i)}} computational complexity of the proposed superposition transmission joint system; \textbf{\textit{ii)}} comparisons with prior works, for example, the joint system based on OFDM and MIMO; \textbf{\textit{iii)}} unifying the performance metric of sensing and communications functions, for example, using the I-MMSE \cite{guo2013interplay} as a bridge; \textbf{\textit{iv)}} scalability to the multiple user (more than two) scenarios; \textbf{\textit{v)}} application of SIC to remove communications signal components at the radar receiving.